\documentclass[a4paper,10pt]{article}
\usepackage{graphicx}
\usepackage{graphics}
\def\F{{\tilde F}}

\title{\huge\bf Two-photon exchange effect on deuteron electromagnetic
form factors}
\author{\bf Yu Bing Dong and D. Y. Chen\\
Institute of High Energy Physics, Chinese Academy of Sciences,\\
Beijing 100049, P. R. China\\
and\\
Theoretical Physics Center for Science Facilities (TPCSF),
CAS, P. R. China\\}
\begin{document}
\maketitle
\begin{abstract}
Corrections of two-photon exchange to proton and neutron electromagnetic
form factors are employed to study the effect of two-photon exchange on
the deuteron electromagnetic form factors. Numerical results of the effect
are given. It is suggested to test the effect in the measurement of $P_z$
in a small angle limit.

\end{abstract}


\section {Introduction}

{\hskip 0.4cm} Proton and deuteron electromagnetic (EM) form factors
have been studied for a long time by unpolarized electron-proton ($ep$) and
electron-deuteron ($eD$) elastic scatterings and by the Rosenbluth separation
\cite {Rosen} based on one-photon-exchange (OPE). The differential cross
section of the $ep$ scattering is
\begin{eqnarray}
d\sigma_0=A_0(\tau_N G_M^2(Q^2)+\epsilon G_E^2(Q^2)),
\end{eqnarray}
with $A_0$ depending on kinematic variables, $\tau_N=Q^2/4M^2_N$ and
$\epsilon =\big [1+2(1+\tau_N)tan^2(\theta/2)\big ]^{-1}$ ($M_N$ and $\theta$
are the nucleon mass and the electron scattering angle). For a long time,
the extracted $Q^2$-dependences of the nucleon EM form factors are believed
to be simple dipole forms. For the proton $G_{E,M}^p$,  one
conventionally gets
\begin{eqnarray}
G^p_E(Q^2)=G^p_M(Q^2)/\mu_p\simeq 1/(1+Q^2(GeV^2)/0.71)^2,
\end{eqnarray}
where $\mu_p=2.79$ is the proton magneton. Recently, the new experiments of
the polarized $ep$ elastic scattering were precisely carried out. The
polarization transfer scattering experiments,
$\vec{e}+p\rightarrow e+\vec{p}$,
show that the ratio $R^p=\mu_pG_E^p(Q^2)/G_M^p(Q^2)\simeq 1-0.158Q^2$
\cite {Jones}. It means that $R^p$ is no longer a simple constant. It
monotonously decreases with the increasing of $Q^2$. This new phenomenon
contradicts to the traditional knowledge shown in eq. (2). \\

One way to resolve this discrepancy is to take the effect of
the two-photon-exchange (TPE) into account \cite
{Gui, Blu, Blu0, Chen, GT, Arr}.
Usually, it is believed that TPE is strongly suppressed by $\alpha_{EM}$
($\alpha_{EM}=1/137$).  However, it was
argued \cite {Gun} that due to the very steep decreasing of the nucleon EM
form factors, the TPE process, where the $Q^2$ is equally shared by the two
exchanging photons, may be compatible to the OPE one. Some calculations of
the TPE corrections to the $ep$ elastic scattering have been done recently
\cite{Gui, Blu, Blu0, Chen, GT, Arr}.
There were also several other works about the TPE effect on the proton charge
radius and on the parity-violating \cite {Car, Blu1} in $ep$ scattering. The
TPE corrections to the deuteron (spin 1 particle) EM form factors and to
the $e^{\pm}p$ processes have been also discussed in Refs. \cite {Dong, Ga1}.
\\

To consider the electron nucleon elastic scattering with
$e(k)+N(p)\rightarrow e(k')+N(p')$ in OPE (${\cal C}=-1$), we have
\begin{eqnarray}
{\cal M}^{el}_{eN}&=&\frac{e^2}{Q^2}
\bar{u}(k')\gamma^{\mu}u(k)\times \bar{u}(p')\Gamma_{\mu}^Nu(p),\nonumber \\
\Gamma^N_{\mu}&=&\Big [\gamma_{\mu}
F_1^N(Q^2)+i\frac{\sigma_{\mu\nu}q^{\nu}}{2M_N} F_2^N(Q^2)
\Big ],
\end{eqnarray}
where $F^N_{1}$ and $F^N_2$ are the conventional Dirac and Pauli form factors
of the nucleon. It should be mentioned that in OPE, the form factors
$F^N_{1,2}$ are real and the functions of $Q^2$ only. \\

If TPE is considered, parity, charge-conjugation and helicity invariances
lead a general expression
\begin{eqnarray}
{\tilde {\cal M}}^{el}_{eN}&=&\frac{e^2}{Q^2}\bigg \{
\bar{u}(k')\gamma_{\mu}u(k)\times\bar{u}(p')\Big [\gamma^{\mu}
\F_1^N(Q^2,\epsilon)+i\frac{\sigma^{\mu\nu}q_{\nu}}{2M_N}
\F_2^N(Q^2,\epsilon)\Big ]u(p)\nonumber \\
&&+\bar{u}(k')\gamma_{\mu}\gamma_5u(k)\times\bar{u}(p')\gamma^{\mu}\gamma^5
\tilde G_A^N(Q^2,\epsilon)u(p)\bigg \}.
\end{eqnarray}
Here, unlike eq. (3), $\F_{1,2}^N(Q^2,\epsilon)$ and
$\tilde G_A^N(Q^2,\epsilon)$ are functions of $Q^2$ and $\epsilon$.
$\F_{1,2}^N(Q^2,\epsilon)$ can be separately expressed in terms of the
contributions of OPE and TPE
\begin{eqnarray}
\F_{1,2}(Q^2,\epsilon)=F_{1,2}(Q^2)+F_{1,2}^{(2)}(Q^2,\epsilon),~~~
\tilde G_{A}(Q^2,\epsilon)=\tilde G_A^{(2)}(Q^2,\epsilon).
\end{eqnarray}
Moreover, according to a general relation \cite {Chen}
\begin{eqnarray}
\frac14\bar{u}(k')\gamma\cdot Pu(k)\times \bar{u}(p')\gamma\cdot Ku(p)&=&
\frac{s-u}{4}\bar{u}(k')\gamma_{\mu}u(k)\times \bar{u}(p')\gamma^{\mu}u(p)
\nonumber \\
&&+\frac{t}{4}\bar{u}(k')\gamma_{\mu}\gamma_5u(k)\times \bar{u}(p')
\gamma^{\mu}\gamma^5u(p),
\end{eqnarray}
where $K=k'+k$, $P=p'+p$, and $s=(p+k)^2$ and $u=(p-k')^2$ are Mandelstam
variables, we have
\begin{eqnarray}
{\tilde {\cal M}}_{eN}^{el}&=&\frac{e^2}{Q^2}
\bar{u}(k')\gamma^{\mu}u(k)\times \bar{u}(p'){\tilde \Gamma}^N_{\mu}u(p),
\nonumber \\
{\tilde \Gamma}^N_{\mu}&=&\bigg \{ \gamma_{\mu}
\F^{'N}_{1}(Q^2,\epsilon)
+\frac{iq^{\nu}\sigma_{\mu\nu}}{2M_N}\F^{'N}_2(Q^2,\epsilon)
+\frac{1}{4M_N^2}\F^{'N}_3(Q^2,\epsilon)\slash\!\!\!{K}P^{\mu}\bigg \}
\end{eqnarray}
where $\tilde F_3^{'N}=\frac{4M_N^2}{t}\tilde G_A^N$ with $t=q^2$ and
\begin{eqnarray}
\F^{'N}_{1}(Q^2,\epsilon)=\F_{1}(Q^2,\epsilon)-\frac{s-u}{4M^2_N}
\F^{'N}_3(Q^2,\epsilon).
\end{eqnarray}
\par\noindent\par
Usually a deuteron is regarded as a weekly bound system of a proton and
a neutron (see Fig. 1). Many calculations for the EM form factors of the
deuteron have been performed in different approaches in the literature
\cite{mathiot,gari,karmanov,kaplan}. Recent calculations based on an
effective Lagrangian approach \cite{Ivanov,Dong0} have shown that one can
reasonably explain the deuteron EM form factors with phenomenological
including two-body operators. Note that the deuteron EM form factors receive
the TPE corrections from many different sources in the effective Lagrangian
approach. For example, the two photons directly couple to one of the nucleons
(see Fig. 2), or the two photons directly couple to one of the two
contact points A and B (the contact points mean the
coupling points of the deuteron and its composite $pn$), There are also
several other interferences between the different OPE couplings. For instant,
one photon couples to A(or B) and another to one of the nucleons, or two
photons respectively couple to the two contact points.\\

In this paper, we'll study the TPE effect on the deuteron EM form factors.
The TPE corrections to the EM form factors of the proton and neutron from
the work of Blunden, Melnitchouk and Tjon \cite {Blu0} will be employed.
We know that in the deuteron EM form factors the contribution from the
direct coupling of the photon to one of the nucleons is more important than
the one from the coupling of a photon to the contact point \cite {Dong0},
and the latter coupling is needed in order to guarantee gauge invariance.
Therefore, it is expected that the TPE effect on the deuteron EM form factors
is dominated by the TPE corrections to the EM form factors of the nucleon
(see Fig. 2), and we, as the first step,  consider the effect of Fig. 2  on
the deuteron EM form factors. This paper is organized as follows. In section
2 the TPE corrections in the $eD$ elastic scattering are briefly discussed.
Numerical results for the corrections to the EM form factors of the deuteron
are displayed in section 3. In Sect. 4, the conclusions will be given.

\begin{figure}[t]
\centering \includegraphics[width=6cm, height=4cm]{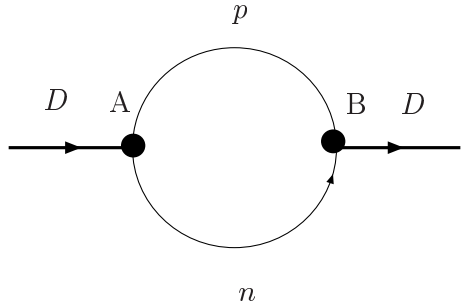}
\caption{\footnotesize Deuteron mass operator}
\end{figure}

\begin{figure}[t]
\centering \includegraphics[width=12cm, height=8cm]{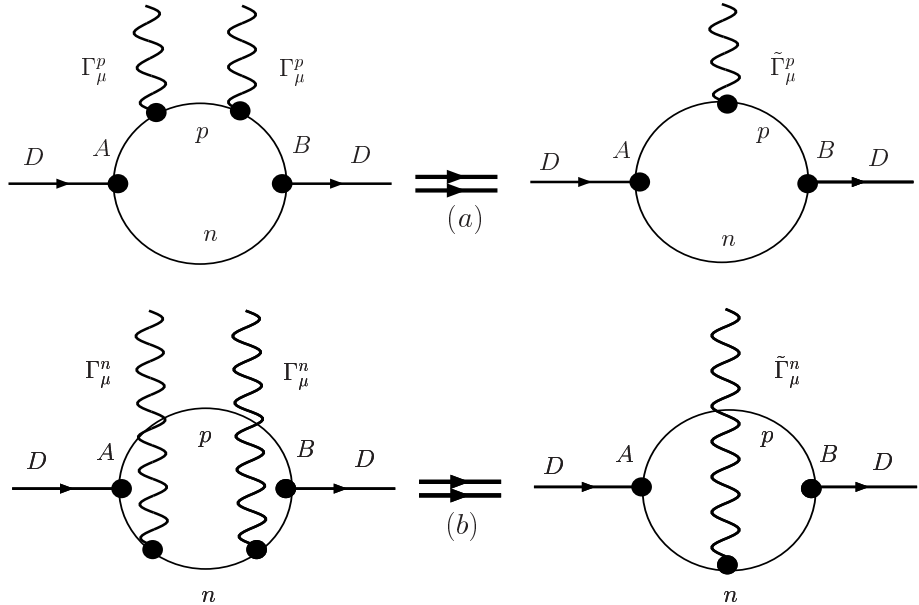}
\caption{\footnotesize Two-photon-exchange effect on deuteron form factors
from the contributions of ${\tilde \Gamma}^p_{\mu}$ (a) and
${\tilde \Gamma}^n_{\mu}$ (b).}
\end{figure}

\section {Two-Photon-Exchange in the $eD$ elastic scattering}

{\hskip 0.4cm}In $eD$ case, the electromagnetic form factors of the deuteron
are defined by the matrix elements of the electromagnetic current $J_\mu(x)$
according to the OPE approximation
\begin{eqnarray}
&&<p_D',~\lambda'\mid J_\mu(0)\mid p_DD,~\lambda> =-e_D\bigg\{\Big
[G_1(Q^2) \xi'^*(\lambda')\cdot \xi(\lambda)\\ \nonumber
&&-G_3(Q^2)\frac{(\xi'^*(\lambda')\cdot q)(\xi(\lambda)\cdot
q)}{2M^2_D}\Big ]\cdot P_\mu
+G_2(Q^2)\Big [\xi_{\mu}(\lambda)(\xi'^*(\lambda ')\cdot q)-
       \xi'^*_{\mu}(\lambda ')(\xi(\lambda)\cdot q)\Big ]\bigg \},
       \label{current}
\end{eqnarray}
where $p'_D,\xi',\lambda'$ (or $p_D,\xi,\lambda $) denote the
momentum, helicity, and polarization vector of the final (or initial)
deuteron, respectively. In eq. (9) $q=p'_D-p_D$ is the photon momentum,
$P=p_D+p'_D, Q^2=-q^2$ is the four-momentum transfer squared, $M_D$
is the deuteron mass, and $e_D$ is the charge of the deuteron. In the
one-photon exchange approximation, the differential cross section of the
unpolarized elastic electron-deuteron scattering
$e(k,s_1)+D(p_D,\xi)\rightarrow e(k',s_3)+D(p'_D,\xi')$
in the laboratory frame is~\cite{Jankus56}
\begin{eqnarray}
\frac{d\sigma}{d\Omega}=\frac{d\sigma}{d\Omega}\bigg
|_{Mott}I_0(OPE), \label{diffCrx}~~~
I_0(OPE)=A(Q^2)+B(Q^2)tan^2\frac{\theta}{2},
\end{eqnarray}
where $\theta$ is the scattering angle of the electron,
$(d\sigma/d\Omega)_{Mott}$ is the Mott cross section for a structure-less
particle with recoil effect, and the two structure functions
\begin{eqnarray}
A(Q^2)=G_C^2(Q^2)+\frac23\tau_D
G_M^2(Q^2)+\frac89\tau^2_DG_Q^2(Q^2),~~~~~~~
B(Q^2)=\frac43\tau_D(1+\tau_D)G^2_M(Q^2). \label{AB}
\end{eqnarray}
In the above eqs. $\tau_D=Q^2/4M^2_D$. $G_M$, $G_C$ and $G_Q$ are the
deuteron magnetic, charge and quadrupole form factors, respectively. They
can be expressed, in terms of $G_1$, $G_2$ and $G_3$, as
\begin{eqnarray}
G_M=G_2, ~~~~G_Q=G_1-G_2+(1+\tau_D)G_3,~~~ G_C=G_1+\frac23\tau_D G_Q.
 \label{GMCQ_G123}
\end{eqnarray}
The normalizations of the three form factors are
$G_c(0)=1,~G_Q(0)=M^2_DQ_D=25.83$, and $G_M(0)=1.714$. Note that in eqs. (10)
and (12), there are two unpolarized structure functions $A$
and $B$, and three independent form factors $G_C$, $G_Q$ and $G_M$ for the
deuteron. To determine the three form factors completely, one needs, at
least, one polarization observable. The optimal choice is the polarization
$T_{20}$ (or $P_{zz}$) \cite {Gar1}. \\

Considering both OPE (${\cal C}=-1$) and TPE (${\cal C}=+1$), and taking
the Lorentz, party, and charge-conjugation invariances into account,
one obtains the most general form of the $eD$ elastic scattering
\cite{Dong, Tarrach75},
\begin{eqnarray}
{\cal M}_{eD}^{el}=\frac{e^2}{Q^2}\bar{u}(k',s_3)
\gamma_{\mu}u(k,s_1)\sum_{i=1}^6G_i'M_i^{\mu},
\label{MeD}
\end{eqnarray}
where
\begin{eqnarray}
M_1^{\mu}&=&(\xi'^*\cdot\xi)P^{\mu},~~~
M_2^{\mu}=\Big [\xi^{\mu}(\xi'^*\cdot q)
-(\xi\cdot q)\xi'^{*\mu}\Big ],\nonumber \\
M_3^{\mu}&=&-\frac{1}{2M_D^2}(\xi\cdot q)(\xi'^*\cdot q)P^{\mu},~~~
M_4^{\mu}=\frac{1}{2M_D^2}(\xi\cdot K)(\xi'^*\cdot K)P^{\mu},\nonumber \\
M_5^{\mu}&=&\Big [\xi^{\mu} (\xi'^*\cdot K)
+(\xi \cdot K)\xi'^{*\mu}\Big ],
\end{eqnarray}
and
\begin{eqnarray}
M_6^{\mu}&=&\frac{1}{2M_D^2}\Big [(\xi\cdot q)(\xi'^*\cdot K)
-(\xi\cdot K)(\xi'^{*}\cdot q)\Big ]P^{\mu}. \label{M1-6}
\end{eqnarray}
General speaking, the form factors $G'_i$
with  $i=1,6$, are complex functions of $s=(p_D+k)^2$ and
$Q^2=-(k-k')^2$. They can be expressed as
\begin{eqnarray}
G_i'(s,Q^2)=G_i(Q^2)+G_i^{(2)}(s,Q^2), \label{Gtilde}
\end{eqnarray}
where $G_i$ represents the contribution arising from the
one-photon exchange, and $G_i^{(2)}$  stands for the rest which
would come mostly from TPE. In the OPE approximation,
$G_4'=G_5'=G_6'=0.$ It is easy to see that $G_i$ $(i=1,2,3)$ are of
order of $(\alpha_{EM})^0$ and $G_i^{(2)}$ ($i=1,...6$) are
of order  $\alpha_{EM}$. Moreover, $G'_i=G_i^{(2)}$ for $i=4,5$ and $6$.\\

To take the TPE corrections to the proton and neutron (see Fig. 2) EM form
factors into
account, and to study the TPE effect on the EM form factors of the deuteron,
we directly calculate the matrix element of
$<p'_D,\lambda'\mid \hat J_{\mu}^p(0)+\hat J_{\mu}^n(0)\mid p_D,\lambda>$,
where
\begin{eqnarray}
\hat J_{\mu}^{p,n}(0)=\mid pn><pn\mid {\tilde \Gamma}_{\mu}^{p,n}(0)
\mid pn><pn\mid.
\end{eqnarray}
The effective interaction between the deuteron and its composites ($pn$)
is \cite{Dong0}
\begin{eqnarray}
{\cal L}_D = g_{D}D^{\mu+}(x)\int dy
\Phi_D(y^2)\bar{p}(x+\frac12y)C\gamma_{\mu}
n(x-\frac12y)+ {\rm H.c.},
\label{ls}
\end{eqnarray}
where $C$ is the charge conjugate matrix. The correlation function
$\Phi_D$ characterizes the finite size of the deuteron as a $pn$ bound state
and depends on the relative Jacobi coordinate $y$, in addition, $x$
being the center-of-mass (CM) coordinate. The Fourier transformation of the
correlation function reads
\begin{eqnarray}
\Phi_D(y^2) \, = \, \int\!\frac{d^4p}{(2\pi)^4}  \,
e^{-ip y} \, \widetilde\Phi_X(-p^2) \,.
\end{eqnarray}
A basic requirement for the choice of an explicit form of the correlation
function is that it vanishes sufficiently fast in the ultraviolet region
of Euclidean space to render the Feynman diagrams ultraviolet finite.
We adopt a Gaussian form,
$\tilde\Phi_D(p_E^2) \doteq \exp( - p_E^2/\Lambda_D^2)\,,$
for the vertex function, where $p_{E}$ is the Euclidean Jacobi momentum.
Here, $\Lambda_D$ is a size parameter, which characterizes the distribution
of the constituents inside the deuteron.\\

The coupling $g_D$ of $<p_D,\lambda \mid pn> =g_D\xi^{'*}(\lambda)$
is determined by the compositeness condition~\cite{Weinberg,Efimov,Anikin,
Faessler}. The condition implies that the renormalization constant of the
hadron wave function is set equal to zero:
\begin{eqnarray}
\label{ZX}
Z_D &=& 1 - \Sigma^\prime_D(M_D^2) = 0.
\end{eqnarray}
Here, $\Sigma^\prime_D(M_{D}^2) = g_{_{D}}^2 \Pi^\prime_D(M_D^2)$ is the
derivative of the transverse part of the mass operator
$\Sigma_D^{\alpha\beta}$, which
conventionally splits into the transverse
$\Sigma_D$ and longitudinal $\Sigma^L_D$  parts as:
\begin{eqnarray}
\Sigma^{\alpha\beta}_{D}(p) = g^{\alpha\beta}_\perp \Sigma_D(p^2)
+ \frac{p^\alpha p^\beta}{p^2} \Sigma^L_D(p^2) \,,
\end{eqnarray}
where
\begin{eqnarray}
g^{\alpha\beta}_\perp = g^{\alpha\beta} - p^\alpha p^\beta/p^2\,,
\hspace*{.2cm} g^{\alpha\beta}_\perp p_\alpha = 0\,.
\end{eqnarray}
The mass operator of the deuteron in our approach is described by
Fig. 1.  If the size parameter $\Lambda_D$ is fixed, the coupling $g_D$
is then obtained according to the compositeness condition eq. (20). \\

Note that the current of photon-nucleon with TPE has an
additional structure ${\tilde F}^{'N}_3$ as shown in eq. (7). An explicit
calculation of the matrix element including TPE (see Fig. 2) gives
\begin{eqnarray}
{\tilde {\cal M}}_{eD}^{el}=\frac{e^2}{Q^2}\bar{u}(k')\gamma_{\mu}u(k)
\xi^{*}_{\sigma}(\lambda')\Big [ {\tilde J}_{p; eD}^{\mu;\sigma\rho}
+{\tilde J}_{n; eD}^{\mu;\sigma\rho}\Big ]
\xi_{\rho}(\lambda),
\end{eqnarray}
where
\begin{eqnarray}
{\tilde J}_{p;eD}^{\mu;\sigma\rho}&=&
g_D^2\int\frac{d^4k}{(2\pi)^4}\Phi_D[(k+\frac{p_D}{2})^2]
\Phi_D[(k+\frac{p'_D}{2})^2]
\nonumber \\
&&\times Tr\bigg (\gamma^{\sigma}S_F(k+p'_D){\tilde \Gamma}^p_{\mu}
S_F(k+p_{D})\gamma^{\rho}S_F(k)\bigg )
\end{eqnarray}
is the contribution from Fig. 2(a) and
\begin{eqnarray}
{\tilde J}_{n;eD}^{\mu;\sigma\rho}&=&
g_D^2\int\frac{d^4k}{(2\pi)^4}\Phi_D[(k-\frac{p_D}{2})^2]
\Phi_D[(k-\frac{p'_D}{2})^2]
\nonumber \\
&&\times Tr\bigg (\gamma^{\sigma}S_F(k)\gamma^{\rho}
S_F(k-p_D){\tilde \Gamma}^n_{\mu}S_F(k-p'_D)\bigg )
\end{eqnarray}
is the one from Fig. 2(b).
From the explicit expressions of eqs. (24-25),
it is found that the form factor of $\F^N_{1,2}$ contributes to the charge,
magnetic and quadrupole form factors of the deuteron. When the TPE effect is
included, it provides a new form factor of the nucleon, ${\tilde F}_3^{'N}$.
This new form factor contributes, in our approach, to the charge and
quadrupole form factors of the deuteron.  Particularly, it also gives a
contribution to $G'_6=G_6^{(2)}$ with the structure of $M_6^{\mu}$
and $q\cdot M_6=0$. The explicit expression for the new form factor of
$G'_6$ contributed by  ${\tilde F}_{3}^{'p}$ is
\begin{eqnarray}
G_6^{\prime p}=-g_D^2\int_0^{\infty}\frac{d\alpha d\beta d\gamma}
{(4\pi)^2\Lambda_D^2Z^2}
e^{-\frac{1}{\Lambda_D^2}[\frac{d_2}{Z}+d_0]} \tilde
F_3^{'p}(Q^2,\epsilon)H_6^p(\alpha,\beta,\gamma,Q^2,\epsilon),
\end{eqnarray}
where
\begin{eqnarray}
H_6^p&=&-4+4x_2-\frac{4\Delta}{M_N^2}+\frac{4\Delta x_2}{M_N^2}
-4\frac{Q^2}{M_N^2}x_1x_2^2\nonumber \\
&&+4\frac{M_D^2}{M_N^2}(-2x_1x_2^2+x_1-x_1^2x_2+x_1^2-x_2^2-x_2^3)
\end{eqnarray}
with
\begin{eqnarray}
Z&=&2+\alpha+\beta+\gamma,~~~~~ \Delta=\frac{\Lambda_D^2}{Z}\nonumber \\
x_1&=&-\frac{\frac12+\beta}{Z},~~~~~
x_2=-\frac{\frac12+\alpha}{Z}\nonumber \\
d_0&=&(\alpha+\beta+\gamma)M_N^2-(\frac12+\alpha+\beta)M_D^2\nonumber \\
d_2&=&(1+\alpha+\beta)^2M_D^2+(\frac12+\alpha)(\frac12+\beta)Q^2,
\end{eqnarray}
and
\begin{eqnarray}
&&\frac{1}{g_D^2}=\frac{1}{8\pi^2}
\int_0^{\infty} \int_0^{\infty} \frac{d\alpha d\beta}{(1+\alpha+\beta)^3}
\exp\biggl( - 2 (\alpha+\beta)\mu_N^2
+ \frac{\alpha+\beta+2\alpha\beta}{2(1+\alpha+\beta)}\mu_D^2
\biggr)\\ \nonumber
&&\times\Bigg( (\alpha+\beta+2\alpha\beta)
\biggl( \mu_N^2+\frac{1}{2 (1+\alpha+\beta)}
+\frac{(1+2\alpha)(1+2\beta)}{4(1+\alpha+\beta)^2}\mu_D^2 \biggr)
+\frac{(1+2\alpha)(1+2\beta)}{2(1+\alpha+\beta)}\Bigg) \nonumber,
\end{eqnarray}
where $\mu_H = m_H/\Lambda_D$ with $H=N, D$. In eqs. (26) and (29),
the integration variables $\alpha,\beta$ and $\gamma$ are the
Feynman parametrizations. The contribution from ${\tilde
F}_{3}^{'n}$ to the new deuteron structure function $G_6^{\prime n}$
can be obtained from eq. (25) in the same way.

\section{Numerical results}

{\hskip 0.4cm}
In \cite {Arr1}, the discrepancies between OPE and TPE have been carefully
analyzed for the $ep$ elastic scattering. The empirical estimates of the
TPE amplitudes in the $ep$ elastic scattering are given based on the
assumptions about the angular dependence of the amplitudes which is limited
by the precision of the Rosenbluth data \cite{Arr2}. It is also assumed that
the entire form factor discrepancies are because of the new form factor
$\tilde F_3^p$.
In a recent paper about a global analysis of the proton elastic form factor
data with the two-photon exchange corrections \cite {Arr3}, the input
TPE corrections are following the formalism of Blunden et al. \cite {Blu0},
rather than the phenomenological corrections extracted  in \cite {Arr1}.
It is found that the value of $Y_{2\gamma}$ is much smaller than that
extracted in the phenomenological analyses. \\

In our numerical estimates of the TPE effect on  the EM form factors of
the deuteron, we will also use the TPE corrections to the proton and
neutron EM form factors following the formalism of Ref. \cite {Blu0}.
It should be mentioned that the TPE corrections are $\theta$-dependent
(or $\epsilon$-dependent). One can get $G_{E,M}^{(2)}/G'_{E,M}$ for the
proton and neutron as well as ${\tilde F}_3^{p,n}(Q^2,\epsilon)$ from the
obtained $Y_{2\gamma}^{p,n}$ with
\begin{eqnarray}
Y_{2\gamma}=Re\Bigg ( \frac{f{\tilde F}_3(Q^2,\epsilon)}{M_N^2\mid
G'_M\mid}\Bigg )
=\frac{K^0}{2M_N}Re\Bigg (\frac{{\tilde
F}_3(Q^2,\epsilon)}{G_M'(Q^2)}\Bigg ),
\end{eqnarray}
where $f=M_N^2\sqrt{\frac{1+\epsilon}{1-\epsilon}}\sqrt{\tau_N(1+\tau_N)}
=\frac12M_NK^0$. To explicitly show the TPE effect on the deuteron EM form
factors, we display, in Figs. 3-5, the ratios $R_i=G_{i}^{(2)}/G'_{i}$ with
$i=C,M,Q$, where $G_i^{(2)}$ are the contributions from TPE of Fig. 2,
and $G'_{i}$ are taken from the phenomenological parametrization of the
deuteron EM form factors \cite {GT1} as empirical data. In the three
figures, we respectively choose the scattering angle $\theta=\pi/6$,
$\pi/2$ and $5\pi/6$. According to the constraint condition that the deuteron
is bound as $<\mid r^{-2}\mid ><0.02GeV^2$ \cite {mathiot}, we select a
typical parameter  $\Lambda_D=0.30~GeV$. Moreover, in Figs. 6 and 7,
we present our predictions for the new form factor $G_6'=G_6^{(2)}$ where
the contributions from Fig. 2(a) and Fig. 2(b) are given, respectively. \\

\begin{figure}[t]
\begin{minipage}[b]{0.5\linewidth}
\centering \includegraphics[width=7cm, height=6cm]{ffigc.eps}
\caption{\footnotesize $R_C(Q^2,\theta)$ for three $\theta$.}
\end{minipage}%
\begin{minipage}[b]{0.5\linewidth}
{\hskip 0.5cm}
\centering \includegraphics[width=7cm, height=6cm]{ffigm.eps}
\caption{\footnotesize $R_M(Q^2,\theta)$ for three $\theta$.}
\end{minipage}
\end{figure}

\begin{figure}[t]
\vspace*{1.0cm}
\centering \includegraphics[width=7cm, height=6cm]{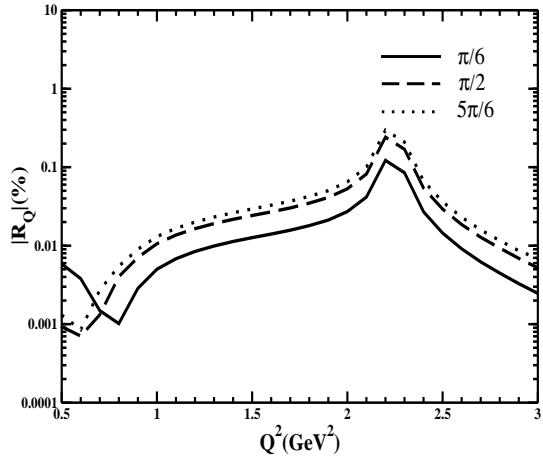}
\caption{\footnotesize $R_Q(Q^2,\theta)$ for three $\theta$.}
\end{figure}

Note that the measured EM form factors of the deuteron appear differently
from those of the nucleons, since they have crossing points. The data
indicate that $G'_C$ or $G'_M$ respectively has a crossing point at
$Q^2\sim 0.7~GeV^2$ or $2.0~GeV^2$ \cite {GT1}. Fig. 5 shows that the
TPE effect on the deuteron quadrupole form factor are greatly reduced
comparing to the corrections to  the proton and neutron.
In Figs. 3 and 4, the peaks result from the crossing points
of $G_C'$ and $G_M'$. The two figures mean that the typical magnitudes of
the ratios $R_C$ and $R_M$, due to the TPE corrections in our approach,
are always less than 10\%, and the corrections to $G_M'$ are more remarkable
than those to $G_C'$. Moreover, Figs. 3-7 tell that the TPE corrections
to $G_{C,M,Q}'$ and to $G_6'$ are $\theta$-dependent. We also find
the two contributions from the proton Fig. 2(a) and neutron Fig. 2(b) always
cancel each other and make the total magnitude of $G_6'$ being smaller. \\

In fact, according to the analyses of Refs. \cite {Dong,Dong0}, it is
expected that
the measurements of the single polarization observables $P_x$  ($T_{11}$) and
$P_z$ ($T_{10}$) are useful to test the TPE effect. Here we know that the
contribution of $G_6'$ to the polarization $P_x$, is
\begin{eqnarray}
P_x^{(2)}\sim
-\frac{4}{3}\frac{K_0}{M_D}
\tau_D\sqrt{\tau_D(1+\tau_D)}tan(\frac{\theta}{2})G_MRe(G'_6).
\end{eqnarray}
Comparing to the $P_x$ in the OPE approximation
\begin{eqnarray}
P_x=-\frac43\sqrt{\tau_D(1+\tau_D)}tan(\frac{\theta}{2})
G_M(G_C+\frac13\tau_D G_Q),
\end{eqnarray}
we find
\begin{eqnarray}
R(P_x)=\frac{P_x^{(2)}}{P_x}=\tau_D\frac{K_0}{M_D}
\frac{Re(G_6')}{G_C+\frac13\tau G_Q}.
\end{eqnarray}
The effect of $G_6'$ on the polarization of $P_x$ is shown in Fig. 8 in
the three cases of $\theta=\pi/6,~\pi/2$ and $5\pi/6$. One sees the ratios are
less than $1\%$ in the range of $0.5\leq Q^2\leq 3~GeV^2$. In Fig. 8 the
maximum points are expected to result from the minimum point of
$G_C+\frac13\tau G_Q$. \\

Moreover, the polarization $P_z$ in OPE is
\begin{eqnarray}
P_z=\frac13\frac{K_0}{M_D}\sqrt{\tau_D(1+\tau_D)}tan^2(\frac{\theta}{2})G_M^2
\end{eqnarray}
and the contribution to $P_z$ from $G_6'$ is
\begin{eqnarray}
P_z^{(2)} =-\frac43\tau_D\sqrt{\tau_D(1+\tau_D)}G_MRe(G_6')
\end{eqnarray}
which is also $\theta$-dependent since $G_6'$ is. In the small angle limit,
the contribution of OPE to $P_z$ vanishes and the one from $G_6'$ remains
no-vanishing to the contrary. Therefore, it is expected that the measurement
of this polarization, $P_z$, in the small $\theta$ limit can easily show
the TPE effect. For the ratio, we get
\begin{eqnarray}
R(P_z)=\frac{P_z^{(2)}}{P_z}=
-4\tau_D\frac{M_D}{K_0}\frac{Re(G'_6)}{tan^2(\frac{\theta}{2})G_M}.
\end{eqnarray}
In Fig. 9, we display the ratio of $R(P_z)$ in the three cases of $\theta$.
A larger TPE effect on $P_z$ than on $P_x$ is seen since the denominator of
the ratio in eq. (36) is proportional to $tan^2(\frac{\theta}{2})$. \\

\begin{figure}[t]
\begin{minipage}[b]{0.5\linewidth}
\vspace*{1.5cm}
\centering
\includegraphics[width=7cm, height=6cm]{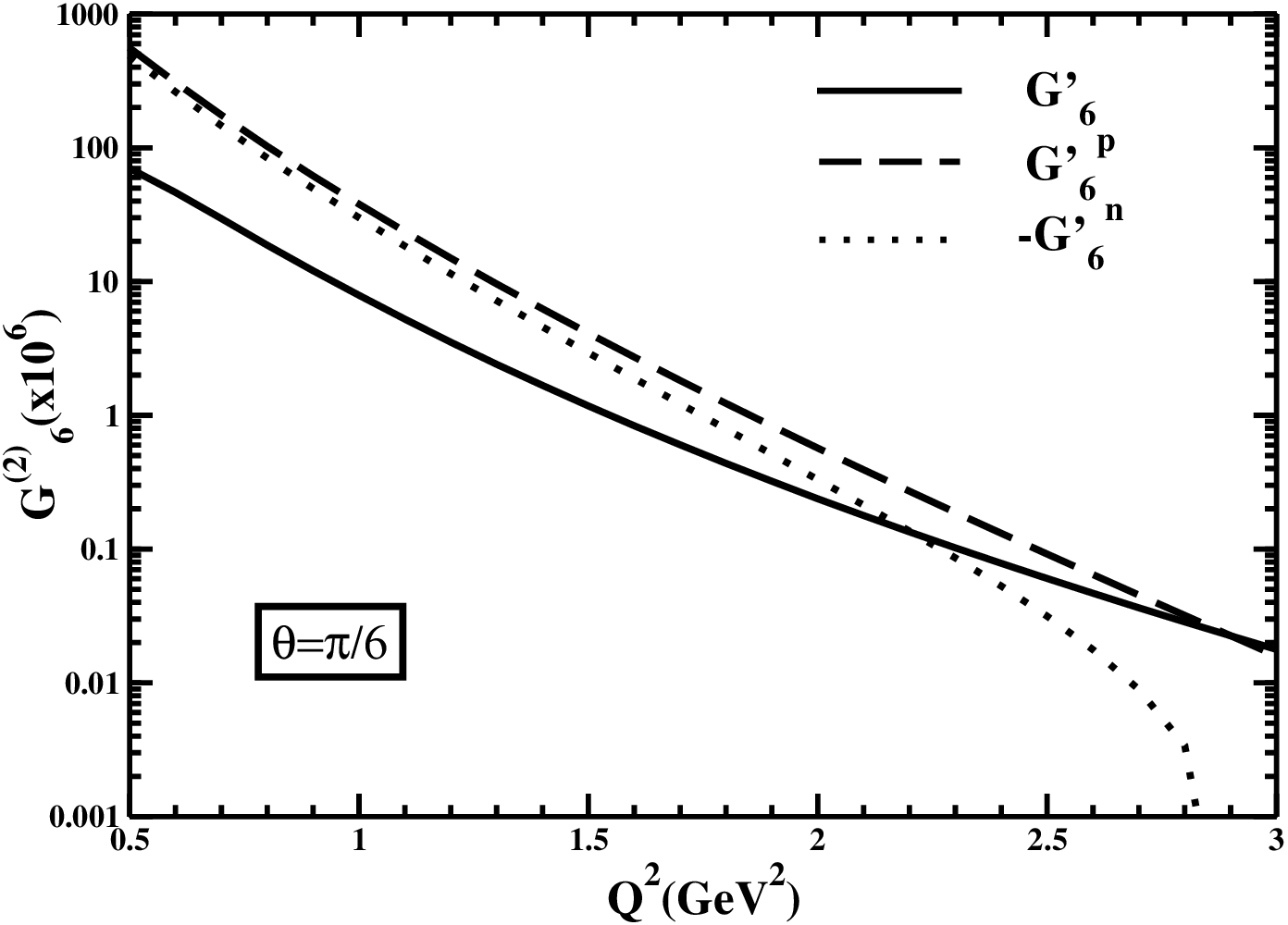}
\caption{\footnotesize $G'_6$ with $\theta=\pi/6$}
\end{minipage}%
{\hskip 0.5cm}
\begin{minipage}[b]{0.5\linewidth}
\centering \includegraphics[width=7cm, height=6cm]{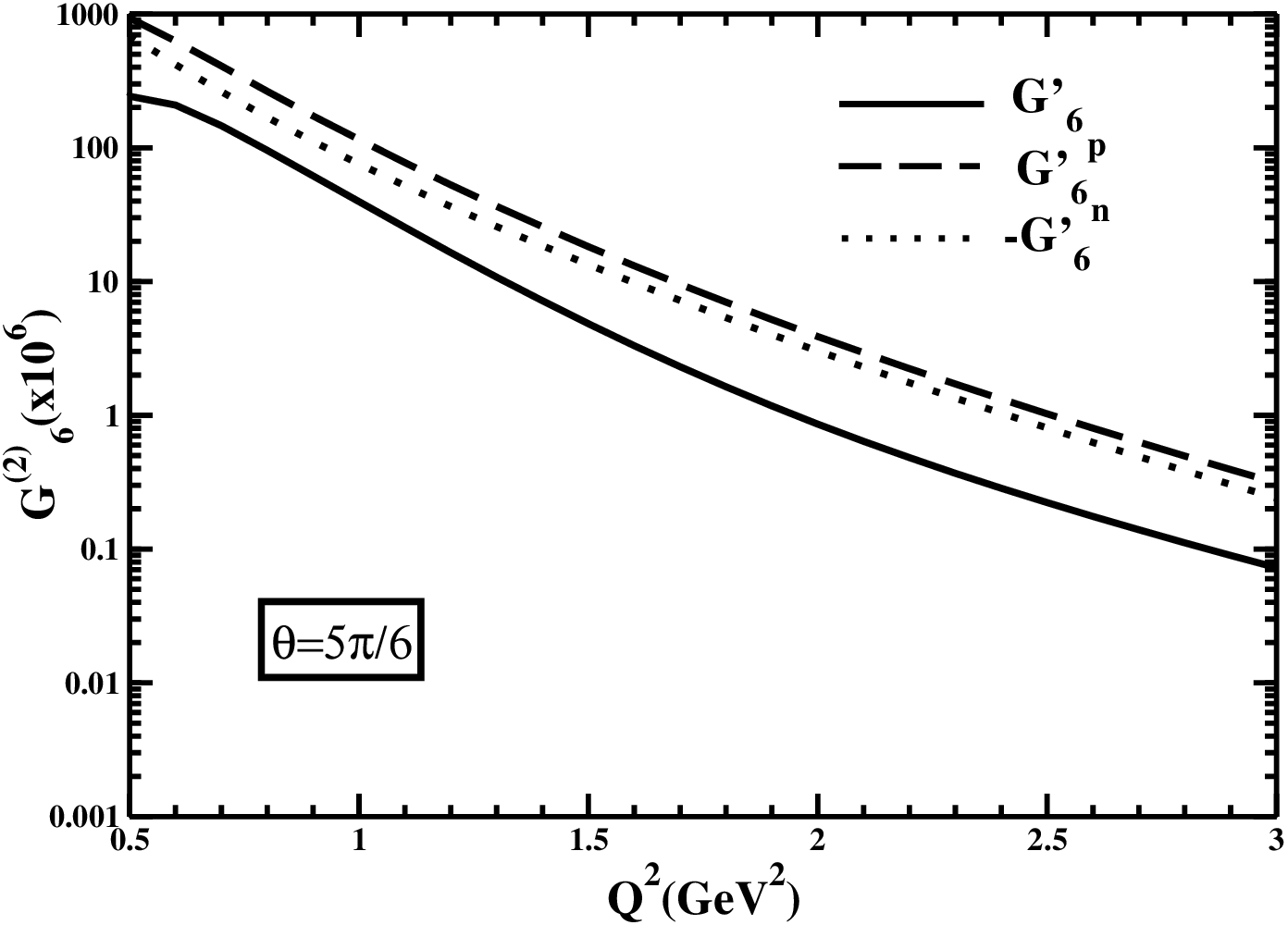}
\caption{\footnotesize $G'_6$ with $\theta=5\pi/6$}
\end{minipage}
\end{figure}

\begin{figure}[t]
\begin{minipage}[b]{0.5\linewidth}
\vspace*{1.5cm}
\centering
\includegraphics[width=7cm, height=6cm]{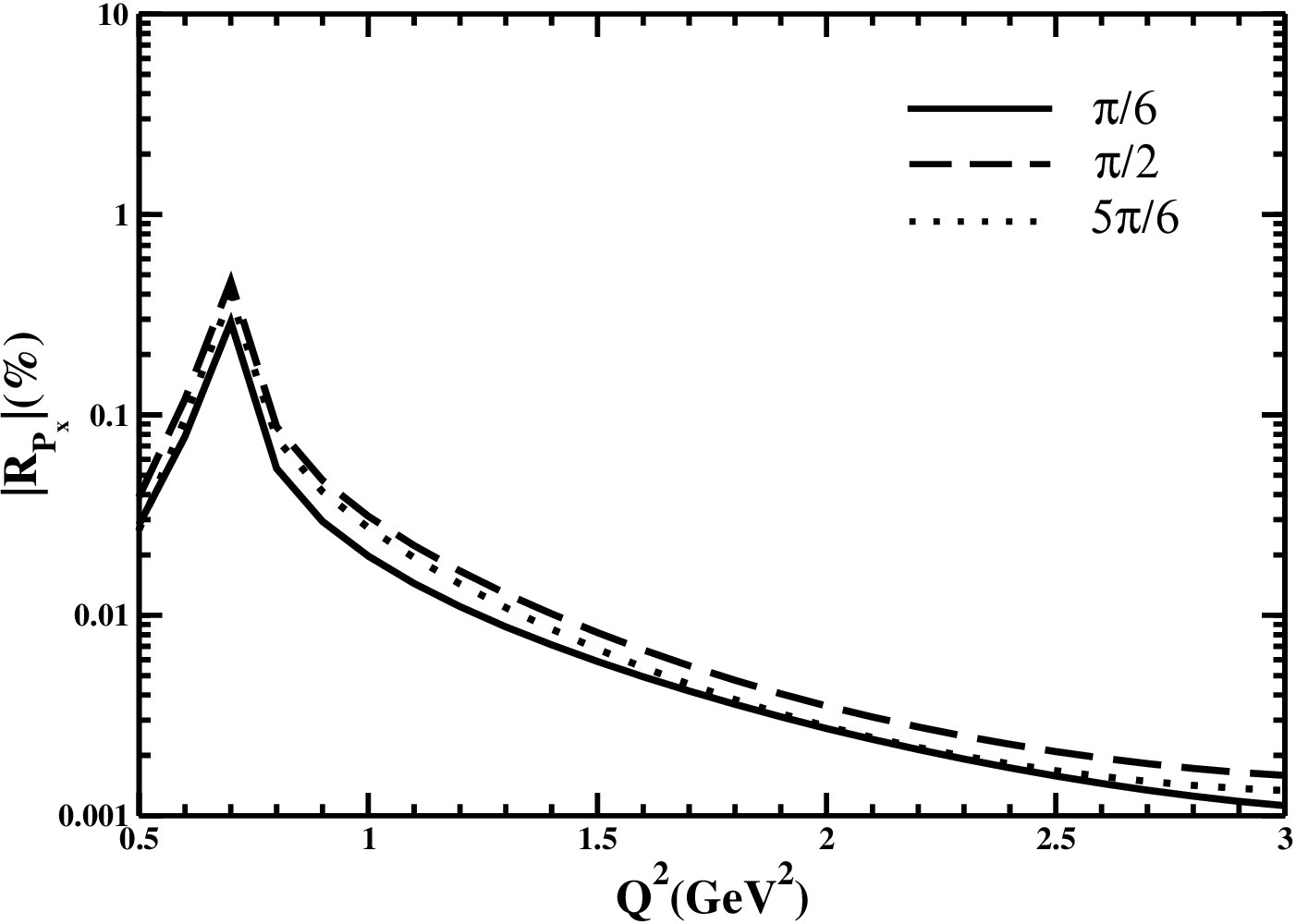}
\caption{\footnotesize $R(P_x)$}
\end{minipage}%
{\hskip 0.5cm}
\begin{minipage}[b]{0.5\linewidth}
\centering \includegraphics[width=7cm, height=6cm]{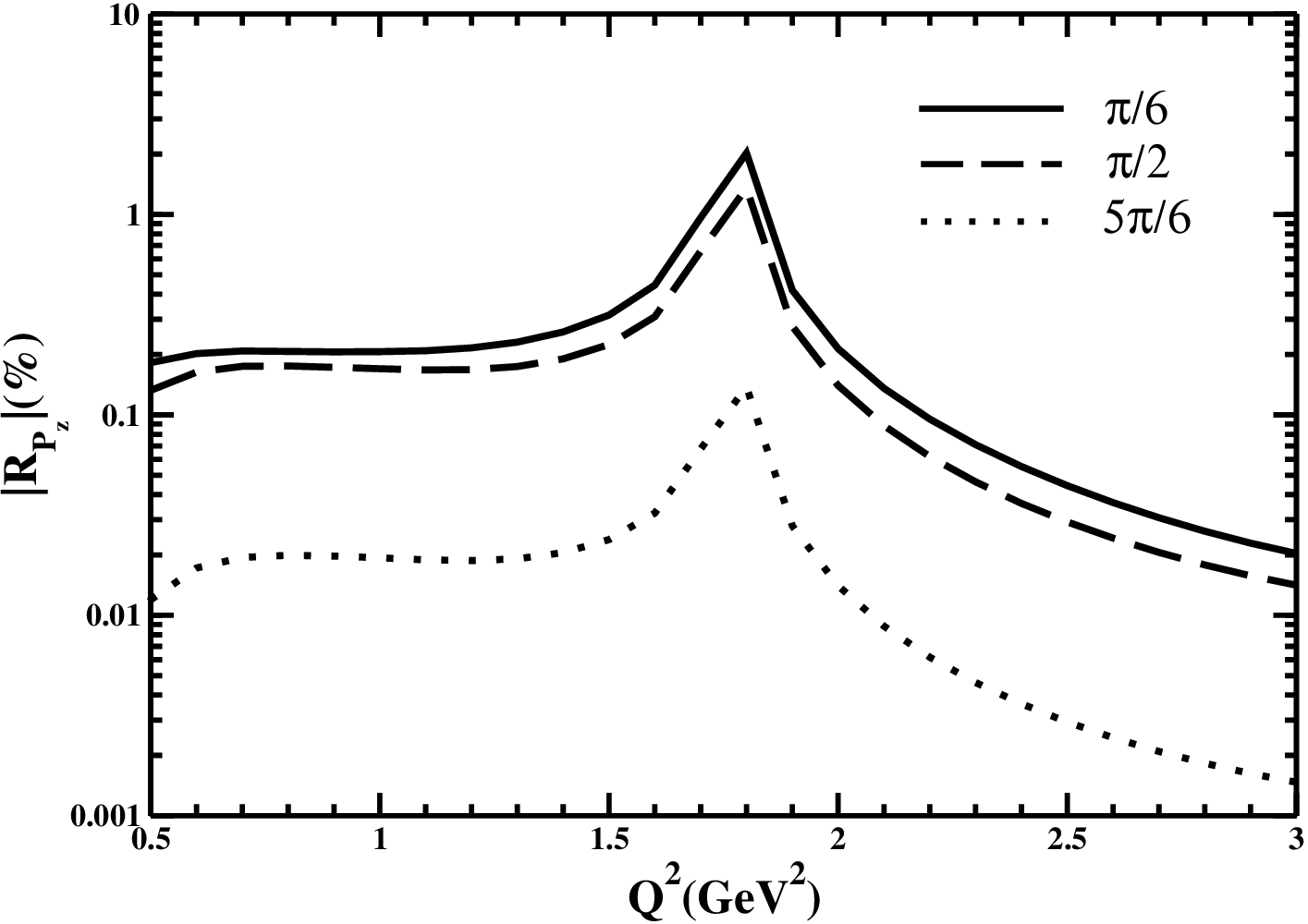}
\caption{\footnotesize $R(P_z)$}
\end{minipage}
\end{figure}

\section {Conclusions}

{\hskip 0.4cm}To summarize, we have explicitly given the TPE corrections
to the conventional form factors of the deuteron $G'_{C,M,Q}$. In our
approach, the TPE corrections to the nucleon EM form factors (see Fig. 2)
are considered. We find that the new from factor of the nucleon with
TPE, $\tilde F'^{p,n}_3$, not only contributes to the form factors
$G_{C,Q}'$ of the deuteron, but also provides a new form factor of the
deuteron $G_6'$. According to the formalism of Ref. \cite {Blu0}, we
numerically estimate the TPE effect on the deuteron
EM form factors, and we get the $\theta$-dependences for all the TPE
corrections. It is suggested the TPE effect can be tested in the measurement
of the single polarization of $P_z$ $(T_{10})$ in the small angle limit.
In addition, we find that the TPE corrections to $G'_M$ are more important
than those to $G'_{C, Q}$. \\

Of course, the TPE effect, we considered in this work, only results from the
sources of the direct couplings of the two photons to one of the two nucleons
inside the deuteron (Fig. 2). There are several other sources of TPE which
could be included in our future calculation as next step. An overall estimate
of all the TPE corrections to the deuteron form factors is in progress.

\section{Acknowledgments}
\par\noindent\par\noindent\par
This work is supported  by the National Sciences Foundations grant
No. 10775148, by CAS grant No. KJCX3-SYW-N2 and in part by the
National Research Council of Thailand through Suranaree University of
Technology and the Commission of High Education, Thailand. Discussions with
S. N. Yang and Valery E. Lyubovitskij are appreciated.

\end{document}